\documentclass[12pt]{article}

\newcommand{\be}{\begin{equation}}
\newcommand{\ee}{\end{equation}}
\newcommand{\bi}[1]{\vspace{-3mm} \bibitem{#1}}

\usepackage{epsfig,amsmath} 
\usepackage{graphics}

\begin{document}

\begin{center}
Annals of Physics, Vol.318. No.2. (2005) pp.286-307
\end{center}

\begin{center}
{\Large \bf Fractional Hydrodynamic Equations for Fractal Media}
\vskip 5 mm

{\large \bf Vasily E. Tarasov } \\

\vskip 3mm

{\it Skobeltsyn Institute of Nuclear Physics, \\
Moscow State University, Moscow 119992, Russia}

{E-mail: tarasov@theory.sinp.msu.ru}
\end{center}
\vskip 11 mm

\begin{abstract}
We use the fractional integrals in order
to describe dynamical processes in the fractal media. 
We consider the "fractional" continuous medium model for the fractal media 
and derive the fractional generalization of the equations of balance of 
mass density, momentum density, and internal energy.
The fractional generalization of Navier-Stokes and Euler 
equations are considered.
We derive the equilibrium equation for fractal media.
The sound waves in the continuous medium model for fractional
media are considered.  
\end{abstract}

PACS:  03.40.Gc; 47.10.+g; 47.53.+n \\ 
%03.40.Gc	Fluid dynamics: general mathematical aspects
%47.10.+g	Fluid dynamics, General theory 
%47.53.+n	Fractals
%47.35.+i	Hydrodynamic waves

Keywords: Hydrodynamic equations; Fractal media; Fractional integrals

%%%%%%%%%%%%%%%%%%%%%%%
\section{Introduction}

The pore space of real media is characterized 
by an extremely complex and irregular geometry \cite{B4,B5,B7}. 
Because the methods of Euclidean geometry, which ordinarily 
deals with regular sets, are purely suited for
describing objects such as in nature, stochastic
models are taken into account \cite{B8}. 
Another possible way of describing a complex structure of 
the pore space is to use fractal theory of sets
of fractional dimensionality \cite{P1}-\cite{SACA}.
Although, the fractal dimensionality does
not reflect completely the geometric and dynamic
properties of the fractal, it nevertheless permits
a number of important conclusions about the behavior 
of fractal structures. For example, if it is
assumed that matter with a constant density is distributed 
over the fractal, then the mass of the fractal enclosed 
in a volume of characteristic dimension
$R$ satisfies the scaling law $ M (R) \sim R^{D}$ , whereas
for a regular n-dimensional Euclidean object $M (R) \sim R^n$.
Let us assume that a network of pore channels
can be treated on a scale $R$ as a stochastic fractal
of dimensionality $D <3$ embedded in a Euclidean
space of dimensionality $n = 3$. Naturally, in real objects 
the fractal structure cannot be observed on all
scales but only those for which $R_p < R < R_0$, where
$R_p$ is the characteristic dimensionality of the pore
channel, and $R_0$ is the macroscopic scale for uniformity 
of the investigated structure and processes.
For example, Katz and Thompson \cite{KT} presented experimental 
evidence indicating that the pore spaces of a set of
sandstone samples are fractals 
%%%and are self-similar over three to four orders of magnitude 
in length extending from 10 angstrom to 100 $\mu m$. 

The natural question arises:
what happens to the "laws of fluid flow"
when the medium through which the fluid actually 
flows is fractional? 
The scientific reason for our interest in this topic 
is immediately obvious: 
the laws associated with fluid flow through fractional 
media are only beginning to be understood \cite{P1}-\cite{Zas2}. 

In the general case, the fractal media cannot 
be considered as continuous media.
There are points and domains that are not filled of the medium particles.
These domains are the porous. 
We suggest to consider the fractal media  as special 
(fractional) continuous media \cite{PLA05}.
We use the procedure of replacement of the fractal medium 
with fractal mass dimension by some continuous medium that 
is described by fractional integrals.
This procedure is a fractional generalization of 
Christensen approach \cite{Chr}.
Suggested procedure leads to the fractional integration and 
differentiation to describe fractal media.
The fractional integrals allow us to take into account 
the fractality of the media.
In order to describe the fractal medium by continuous medium model
we must use the fractional integrals. 

In many problems the real fractal structure of matter 
can be disregarded and the medium can be replaced by  
some "fractional" continuous mathematical model. 
In order to describe the medium with 
non-integer mass dimension, we must use the fractional calculus.
Smoothing of the microscopic characteristics over the 
physically infinitesimal volume, we transform the initial 
fractal medium into "fractional" continuous model
that uses the fractional integrals. 
The order of fractional integral is equal 
to the fractal mass dimension of the medium.

More consistent approach to describe the fractal media
is connected with the mathematical definition the integrals
on fractals. In \cite{RLWQ}, was proved that integrals 
on net of fractals can be approximated by fractional integrals. 
In \cite{chaos}, we proved that fractional integrals 
can be considered as integrals over the space with fractional 
dimension up to numerical factor. To prove we use the well-known 
formulas of dimensional regularizations \cite{Col}.  

The fractional continuous models of fractal media
can have a wide application. 
This is due in part to the relatively small numbers of parameters 
that define a random fractal medium of great complexity
and rich structure. The fractional continuous model allows us
to describe dynamics of wide class of fractal media.  
Fractional integrals can be used to derive the fractional 
generalization of the equations of balance for the fractal media.

In this paper, we use the fractional integrals in order
to describe dynamical processes in the fractal media. 
In section 2,  we consider the "fractional" continuous medium model 
for the fractal medium.
In section 3-5, we derive the fractional generalization of 
the equations of balance of 
mass density, momentum density, and internal energy.
In section 6, the fractional generalization of Navier-Stokes and Euler 
equations are considered.
In section 7, we derive the equilibrium equation for fractal media.
In section 8, we consider the fractional generalization of Bernoulli integral.
In section 9, the sound waves in the continuous medium model for fractional
media are considered.  
Finally, a short conclusion is given in section 10.

%%%%%%%%%%%%%%%%%%%%%%%%%%%%%%%%%%%%%%%%%%%
\section{Fractal Media and Fractional Integrals}

The cornerstone of fractals is the meaning of dimension, 
specifically the fractal dimension. 
Fractal dimension can be best calculated by box counting 
method which means drawing a box of size $R$ 
and counting the mass inside. 
The mass fractal dimension \cite{Mand,Schr} can be easy measured
for fractal media.

The properties of the fractal media like mass obeys a power law relation
\be \label{MR} M(R) =kR^{D} , \quad (D<3),  \ee
where $M$ is the mass of fractal medium, $R$ is a box size (or a sphere radius),
and $D$ is a mass fractal dimension. 
Amount of mass of a medium inside a box of size $R$
has a power law relation (\ref{MR}).

The power law relation (\ref{MR}) can be naturally 
derived by using the fractional integral.
In this section, we prove that the mass fractal dimension 
is connected with the order of fractional integrals. 
Therefore, the fractional integrals can be used to describe fractal
media with non-integer mass dimensions. 

Let us consider the region $W_A$ in three-dimensional 
Euclidean space $E^3$, where $A$ is the midpoint of this region.
The volume of the region $W_A$ is denoted by $V(W_A)$.
If the region $W_A$ is a ball with the radius $R_A$,
then the midpoint $A$ is a center of the ball, and
the volume $V(W_A)=(4/3)\pi R^3_A$ .
The mass of the region $W_A$ in the fractal medium is denoted 
by $M_D(W_A)$, where $D$ is a mass dimension of the medium. 

The fractality of medium means that the mass of this medium 
in any region $W_A$ of Euclidean space $E^3$ increase more slowly 
that the volume of this region.
For the ball region of the fractal media, 
this property can be described by the power law (\ref{MR}),
where $R$ is the radius of the ball $W_A$ that is 
much more than the mean radius $R_p$ of the porous sphere.

Fractal media are called homogeneous fractal media if the power 
law (\ref{MR}) does not depends on the translation and 
rotation of the region. The homogeneity property of the media
can be formulated in the form:
%%%\noindent {\bf Homogeneity}: 
For all regions $W_A$ and $W_B$ of the homogeneous fractal media 
such that the volumes are equal $V(W_A)=V(W_B)$, 
we have that the mass of these regions are equal $M_D(W_A)=M_D(W_B)$. 
Note that the wide class of the fractal media satisfies 
the homogeneous property.
In many cases, we can consider the porous media \cite{Por1,Por2}, 
polymers \cite{P}, colloid agregates \cite{CA}, and 
aerogels \cite{aero} as homogeneous fractal media.

To describe the fractal medium, we must use the 
continuous medium model such that the fractality 
and homogeneity properties can be realized in the form: \\

\noindent
(1) Fractality:
The mass of the ball region $W$ of fractal medium obeys a power law relation
\be \label{MR2} M_D(W)=M_0 \Bigl( \frac{R}{R_p} \Bigr)^D , \ee
where $D<3$ and $R$ is the radius of the ball. 
In the general case, we have the scaling law relation
\[ dM_D(\lambda W)=\lambda^D dM_D(W) , \]
where $\lambda W=\{\lambda x, \ \ x \in W \}$. \\

\noindent
(2) Homogeneity:
The local density of homogeneous fractal medium is
translation invariant value that have the form
$\rho({\bf r})=\rho_0=const$.

\vskip 3mm

We can realize these requirements by the 
fractional generalization of the equation
\be \label{MW} M_3(W)=\int_W \rho({\bf r}) d^3 {\bf r} . \ee
Let us define the fractional integral
in Euclidean space $E^3$ in the Riesz form
\cite{SKM} by the equation
\be \label{ID} (I^{D}\rho)({\bf r}_0)=
\int_W \rho({\bf r}) dV_D , \ee
where $dV_D=c_3(D,r,r_0)d^3 {\bf r}$, and 
\[ c_3(D,r,r_0)=\frac{2^{3-D} \Gamma(3/2)}{\Gamma(D/2)} 
|{\bf r}-{\bf r}_0|^{D-3} ,  \quad
|{\bf r}-{\bf r}_0|=\sqrt{\sum^3_{k=1} (x_k-x_{k0})^2}. \]
The point ${\bf r}_0 \in W$ is the initial point of the fractional integral.
We will use the initial points in the integrals are set to zero (${\bf r}_0=0$).
The numerical factor in Eq. (\ref{ID}) has this form in order to
derive usual integral in the limit $D\rightarrow (3-0)$.
Note that the usual numerical factor
$\gamma^{-1}_3(D)={\Gamma(1/2)}/{2^D \pi^{3/2} \Gamma(D/2)}$,
which is used in Ref. \cite{SKM},  
leads to $\gamma^{-1}_3(3-0)= {\Gamma(1/2)}/{2^3 \pi^{3/2} \Gamma(3/2)}$ 
in the limit $D\rightarrow (3-0)$. 

Using notations (\ref{ID}), we can rewrite Eq. (\ref{MW})
in the form $M_3(W)=(I^{3}\rho)({\bf r}_0)$. 
Therefore the fractional generalization of this equation can be
defined in the form
\be \label{MWD}  M_D(W)=(I^D \rho)({\bf r}_0)=
 \frac{2^{3-D} \Gamma(3/2)}{\Gamma(D/2)}
\int_W \rho({\bf r}) |{\bf r}-{\bf r_0}|^{D-3} d^3 {\bf r} . \ee
If we consider the homogeneous fractal media 
($\rho({\bf r})=\rho_0=const$) and the ball region $W$, then
we have 
\[ M_D(W)= \rho_0 \frac{2^{3-D} \Gamma(3/2)}{\Gamma(D/2)} 
\int_W |{\bf R}|^{D-3} d^3 {\bf R} . \]
where ${\bf R}={\bf r}-{\bf r}_0$. 
Using the spherical coordinates, we get
\[ M_D(W)= \frac{\pi 2^{5-D} \Gamma(3/2)}{\Gamma(D/2)} 
\rho_0 \int_W R^{D-1} d R= 
\frac{2^{5-D} \pi \Gamma(3/2)}{D \Gamma(D/2)} \rho_0 R^{D} , \]
where $R=|{\bf R}|$. 
As the result, we have $M(W)\sim R^D$, i.e., we derive Eq. (\ref{MR2})
up to the numerical factor.
Therefore, the fractal medium with non-integer mass dimension $D$ can be
described by fractional integral of order $D$.

Note that the interpretation of the fractional integration
is connected with fractional dimension \cite{chaos}.
This interpretation follows from
the well-known formulas for dimensional regularizations \cite{Col}:
\be \label{dr} \int f(x) d^{D} x =
\frac{2 \pi^{D/2}}{\Gamma(D/2)}
\int^{\infty}_{0} f(x)  x^{D-1} dx  . \ee
Using Eq. (\ref{dr}), we get
that the fractional integral 
\[ \int_W f(x) dV_D, \]
can be considered as a 
integral in the fractional dimension space
\be \label{fnc-2} \frac{\Gamma(D/2) }{2 \pi^{D/2} \Gamma(D)}
\int  f(x) d^{D} x  \ee
up to the numerical factor
$\Gamma(D/2) /( 2 \pi^{D/2} \Gamma(D))$.

%%%%%%%%%%%%%%%%%%%%%%%%%%%%%%%%%%%%%%%%%%%%%%%%%%%%%%%%%%%%%%%%%%
%%%\section{Fractional Equation of Continuity}
%%%%%%%%%%%%%%%%%%%%%%%%%%%%%%%%%%%%%%%%%%%%%%%%%%%
\section{Equation of Balance of Mass Density}

The fractional integrals can be used not only to calculate 
the mass dimensions of fractal media. 
Fractional integration can be used to describe 
the dynamical processes in the fractal media.
Using fractional integrals, we can derive the fractional generalization 
of dynamical equations \cite{chaos,PRE05}.  
In this section, we derive the fractional analog of the equation 
of continuity for the fractal media.

Let us consider the region $W$ of the medium. 
The boundary of this region is denoted by $\partial W$.
Suppose the medium in the region $W$ has the mass dimension $D$.
In general, the medium on the boundary $\partial W$ has the dimension $d$. 
In the general case, the dimension $d$ is not equal to $2$ and 
is not equal to $(D-1)$. 

The balance of the mass density is described by the equation
\be \label{BM} \frac{d}{dt} \int_W \rho({\bf R},t) dV_D=0, \ee 
where we use $d M_D(W)/dt=0$ and the following notations:
\[ dV_D=\frac{2^{3-D} \Gamma(3/2)}{\Gamma(D/2)} 
|{\bf R}|^{D-3} dV_3 , \quad d V_3=d^3 {\bf R} . \]
Here, and late we use the initial points in the integrals 
are set to zero (${\bf r}_0=0$).
The integral (\ref{BM}) is considered for the region $W$ 
which moves with the medium. 
The field of the velocity is denoted by ${\bf u}={\bf u}({\bf R},t)$. 
The total time derivative of the volume integral is defined by the equation
\be \label{ttd11} \frac{d}{dt} \int_W A dV_D=
\int_W \frac{\partial A}{\partial t} dV_D+
\int_{\partial W} A u_n dS_d . \ee
Here, $u_n$ is defined by $u_n=({\bf u},{\bf n})=u_kn_k$, 
the vector ${\bf u}=u_k{\bf e}_k$ is a velocity field, 
and ${\bf n}=n_k {\bf e}_k$ is a vector of normal. 
The surface integral for the boundary $\partial W$ can be represented as
a volume integral for the region $W$.
To realize the representation, we derive the fractional 
generalization of the Gauss theorem (see Appendix). 
As the result, we have the following equation for the total time 
derivative (\ref{ttd11}) of the volume integral
\be \label{FTTD}
\frac{d}{dt} \int_W A dV_D=\int_W \Bigl( \frac{\partial A}{\partial t}+
c^{-1}_3(D,R) div( c_2(d,R) A {\bf u})\Bigr)  dV_D . \ee 
For the integer dimensions $d=2$ and $D=3$, we have the usual equation.  
Let us introduce notations that simplify the form of equations.
We use the following generalization of the total time derivative
\be
\Bigl(\frac{d}{dt}\Bigr)_D=\frac{\partial}{\partial t}+
c(D,d,R) u_k \frac{\partial}{\partial x_k} ,
\ee
where the coefficient $c(D,d,R)$ is defined by 
\[ c(D,d,R)=c^{-1}_3(D,R) c_2(d,R)=
\frac{2^{D-d-1} \Gamma(D/2)}{\Gamma(3/2) \Gamma(d/2)} |{\bf R}|^{d+1-D} . \]
Note that the media with integer dimensions ($D=3$, $d=2$) have $c(D,d,R)=1$. 
We use the following generalization of the divergence
\be \label{DivD}
Div_D ( {\bf u})= c^{-1}_3(D,R) \frac{\partial }{\partial {\bf R}} 
\Bigl( c_2(d,R) {\bf u} \Bigr)=
\frac{2^{D-d-1} \Gamma(D/2)}{\Gamma(3/2) \Gamma(d/2)} 
|{\bf R}|^{3-D} div( |{\bf R}|^{d-2} {\bf u}) , \ee
and the derivative with respect to the coordinates
\be \label{nablaD}
\nabla^D_k A= c^{-1}_3(D,R) \frac{\partial c_2(d,R) A}{\partial x_k} =
\frac{2^{D-d-1} \Gamma(D/2)}{\Gamma(3/2) \Gamma(d/2)} 
|{\bf R}|^{3-D} \frac{\partial }{\partial x_k}  |{\bf R}|^{d-2} A . \ee
Here, $Div_D ={\bf e}_k \nabla^D_k$ and $Div_D(A {\bf u})=\nabla^D_k (A u_k)$. 
Note that the rule of term-by-term differentiation 
for the operator $\nabla^D_k$ is not satisfied
\[ \nabla^D_k(AB)\not= A\nabla^D_k(B)+B\nabla^D_k(A) . \]
This operator satisfies the following rule:
\be \label{7} 
\nabla^D_k (AB)=A\nabla^D_k (B)+c(D,d,R) B 
\frac{\partial A}{\partial x_k} . \ee 
Note that $\nabla^D_k (1)\not=0$, and we have
\[ \nabla^D_k (1)=c(D,d,R) (d-2) x_k/R^2 . \]
Using these notations, we rewrite Eq. (\ref{FTTD}) 
for the total time derivative of the integral in an equivalent form
\be \label{FTTD2}
\frac{d}{dt} \int_W A dV_D=\int_W \Bigl( \Bigl(\frac{d}{dt}\Bigr)_D A+
A \ Div_D({\bf u})\Bigr) dV_D . \ee 

To derive the fractional generalization of the equation
of continuity \cite{chaos}, we consider $A=\rho({\bf R},t)$ in Eq. (\ref{FTTD2})
and the equation of balance of mass (\ref{BM}).
Substituting $A=\rho({\bf R},t)$ in Eq. (\ref{FTTD2}), we get
\be \frac{d}{dt} \int_W \rho dV_D=\int_W \Bigl( 
\Bigl(\frac{d}{dt}\Bigr)_D \rho+\rho Div_D({\bf u}) \Bigr)  dV_D . \ee 
Therefore, equation of balance of mass (\ref{BM}) has the form
\[ \int_W \Bigl( 
\Bigl(\frac{d}{dt}\Bigr)_D \rho+\rho Div_D({\bf u}) \Bigr)  dV_D=0. \]
This equation is satisfied for all regions $W$.
Therefore, we have the fractional equation of continuity
\be \Bigl(\frac{d}{dt}\Bigr)_D \rho+\rho Div_D({\bf u})=0 . \ee
The fractional generalization of equation of continuity
has the form
\[  \Bigl(\frac{d}{dt}\Bigr)_D \rho=-\rho \nabla^D_k u_k . \]
Using Eqs. (\ref{c2}) and (\ref{c3}) of Appendix, 
the equation of continuity can rewritten in an equivalent form
\be \frac{\partial \rho}{\partial t}+ 
\frac{2^{D-d-1} \Gamma(D/2)}{\Gamma(3/2) \Gamma(d/2)} |{\bf R}|^{3-D}
div(  |{\bf R}|^{d-2} \rho {\bf u})=0 . \ee
Using
\[ grad |{\bf R}|^{d-2}= \frac{\partial |{\bf R}|^{d-2}}{\partial {\bf R}}=
(d-2)|{\bf R}|^{d-3} \frac{\partial |{\bf R}|}{\partial {\bf R}}=
(d-2)|{\bf R}|^{d-4} {\bf R} , \]
we have the equation of continuity in the form
\be \frac{\partial \rho}{\partial t}+ c(D,d,R) {\bf u} 
\frac{\partial \rho}{\partial {\bf R}}+c(D,d,R) \rho \Bigl( div ({\bf u})+ 
(d-2)\frac{({\bf u},{\bf R})}{|{\bf R}|^2} \Bigr)=0 , \ee
where we use the following notation $({\bf u},{\bf R})= u_k x_k$. 

For the homogeneous media, we have $\rho=const$ and 
the equation of continuity leads us to the equation
\[ div ({\bf u})+\frac{(d-2) ({\bf R},{\bf u})}{|{\bf R}|^2}=0 . \]
Therefore, we get the non-solenoidal field of the velocity 
($div {\bf u}=\partial {\bf u}/ \partial {\bf R} \not=0$).

In addition to mass density $\rho$, the continuity equation includes 
the density of momentum $\rho {\bf u}$. 
To obtain the equation for the density of momentum,
we consider the mass force and surface force.

%%%%%%%%%%%%%%%%%%%%%%%%%%%%%%%%%%%%%%%%%%%%%%%%%%%
\section{Equation of Balance of Momentum Density}

Let the force ${\bf f}=f_k {\bf e}_k$ be a function of 
the space-time point $({\bf R},t)$.
The force ${\bf F}^M$, that acts on the mass $M_D(W)$ of the 
medium region $W$, is defined by 
\be \label{FM} {\bf F}^M =\int_W \rho({\bf R},t) {\bf f}({\bf R},t) dV_D.  \ee
The force ${\bf F}^S$, that acts on the surface of the boundary 
$\partial W$ of continuous medium region $W$, is defined by 
\be \label{FS} {\bf F}^S=\int_{\partial W} {\bf p}_n({\bf R},t) dS_{d},  \ee
where ${\bf p}={\bf p}({\bf R},t)$ is a density of 
the surface force, and
${\bf p}_n=p_{kl} n_k {\bf e}_k$.
Here, ${\bf n}=n_k {\bf e}_k$ is the vector of normal. 

Let ${\bf P}$ be a momentum of the medium mass that is situated 
in the region $W$. If the mass $dM_D(W)=\rho({\bf R},t) dV_D$ moves with
the velocity ${\bf u}$, than the momentum of this mass is 
$d{\bf P}=dM_D(W) {\bf u}=\rho {\bf u} dV_D$. 
The momentum ${\bf P}$ of the mass of the region $W$ is defined by
the equation
\be \label{P} 
{\bf P}=\int_W \rho({\bf R},t) {\bf u}({\bf R},t) dV_D . \ee
The equation of balance of density of momentum
\be \label{ML1} \frac{d {\bf P}}{dt}={\bf F}^M+{\bf F}^S . \ee
Substituting Eqs. (\ref{FM}) - (\ref{P}) into Eq. (\ref{ML1}),
we get the balance equation in the form
\be \label{ML2} \frac{d}{dt}\int_W \rho({\bf R},t) {\bf u}({\bf R},t) dV_D=
\int_W \rho({\bf R},t) {\bf f}({\bf R},t) dV_D+
\int_{\partial W} {\bf p}_n({\bf R},t) dS_{d} . \ee 
Using Eq. (\ref{FGT}) of Appendix, the surface integral can be 
represented as the following volume integral
\[ \int_{\partial W} {\bf p}_n dS_{d} =
\int_{\partial W} c_2(d,R) {\bf p}_n dS_{2} 
=\int_W  \frac{\partial 
(c_2(d,R) {\bf p}_l)}{\partial x_l} c^{-1}_3(D,R) dV_D=
\int_W  \nabla^D_l {\bf p}_l dV_D . \]
The balance equation of momentum density can be
written for the components of vectors ${\bf u}=u_k {\bf e}_k$, \ 
${\bf f}=f_k {\bf e}_k$,  and ${\bf p}_l=p_{kl} {\bf e}_k$, in the form
\[ \frac{d}{dt}\int_W \rho u_k dV_D=\int_W \Bigl( \rho f_k +
\nabla^D_l p_{kl} \Bigr) dV_D . \]
The relation for the derivative of the volume integral 
with respect to time has the form (\ref{FTTD2}). 
Using this relation for $A=\rho u_k$, we get 
the equation for the total time derivative of the integral 
\be 
\frac{d}{dt} \int_W \rho u_k dV_D=
\int_W \Bigl( \Bigl(\frac{d}{dt}\Bigr)_D (\rho u_k)+
(\rho u_k) \ Div_D({\bf u})\Bigr)  dV_D . \ee 
Therefore, we can rewrite 
the balance momentum density (\ref{ML2}) in the form
\[ \int_W \Bigl( \Bigl(\frac{d}{dt}\Bigr)_D (\rho u_k)+
(\rho u_k) \ Div_D({\bf u})
-\rho f_k -  \nabla^D_l p_{kl}  \Bigr) dV_D=0 . \]
This equation is satisfied for all regions $W$. Therefore, 
\[ \Bigl(\frac{d}{dt}\Bigr)_D (\rho u_k)+
(\rho u_k) \ Div_D({\bf u}) 
-\rho f_k -  \nabla^D_l p_{kl} =0 . \]
Using the rule of the term-by-term differentiation with respect to time
\[ \Bigl(\frac{d}{dt}\Bigr)_D(\rho u_k)=\rho \Bigl(\frac{d}{dt}\Bigr)_D u_k+
u_k \Bigl(\frac{d}{dt}\Bigr)_D \rho , \]
we get the following form of the equations
\[ \rho \Bigl(\frac{d}{dt}\Bigr)_D u_k+
u_k \Bigl(\Bigl(\frac{d}{dt}\Bigr)_D \rho+\rho \ Div_D ({\bf u}) \Bigr)-
\rho f_k- \nabla^D_l p_{kl}=0 . \]
Using the continuity equation, we reduce the fractional equation 
of balance of density of momentum to the form
\be \label{EM} 
\rho \Bigl(\frac{d}{dt}\Bigr)_D u_k=
\rho f_k+\nabla^D_l p_{kl} . \ee
These equations can be called the equation of balance of momentum 
of fractal medium.

%%%%%%%%%%%%%%%%%%%%%%%%%%%%%%%%%%%%%%%%%%%%%%%%%%%
\section{Equation of Balance of Energy Density}

In the general case, the internal energy for the inhomogeneous medium 
is a function of the space-time point $({\bf R},t)$: $e=e({\bf R},t)$.
The internal energy $dE$ of the mass $dM_D(W)$ is equal to
$dE=e({\bf R},t)\rho ({\bf R},t) dV_D$.
The internal energy of the mass of the region $W$ is defined
by the equation
\[ E=\int_W \rho({\bf R},t) e({\bf R},t) dV_D . \]
The kinetic energy $dT$ of the mass $dM_D(W)=\rho d V_D$, which moves with
the velocity ${\bf u}={\bf u}({\bf R},t)$, is equal to
\[ dT=dM_D \frac{{\bf u}^2}{2}=\frac{\rho {\bf u}^2}{2} dV_D .\] 
The kinetic energy of the mass of the region $W$ is
\[ T=\int_W \frac{\rho {\bf u}^2}{2}  dV_D . \]
The total energy is a sum of the kinetic and internal energies 
\[ U=T+E=\int_W \rho\Bigl( \frac{{\bf u}^2}{2}+e\Bigr) dV_D .\]
The change of the total energy is defined by
\[ U(t_2)-U(t_1)=A_{M}+A_S+Q_S , \]
where $A_M$ is the work of mass forces; $A_M$ is the work of surface forces; 
$Q_S$ is the heat that are influx into the region.  

The mass $dM_D(W)=\rho dV_D$ is subjected to force ${\bf f} \rho dV_D$.
The work of this force is $({\bf u},{\bf f}) \rho dV_D dt$, where 
$({\bf u},{\bf f})= u_k f_k$. 
The work of the mass forces for the region $W$ and time interval $[t_1;t_2]$ 
is defined by the following equation
\[ A_M=\int^{t_2}_{t_1} dt \int_W ({\bf u},{\bf f}) \rho dV_D . \]
The surface element $dS_d$ is subjected to force ${\bf p}_n dS_d$.
The work of this force is $({\bf p}_n,{\bf u}) dS_{d} dt$. 
The work of the surface forces for the region $W$ and time interval $[t_1;t_2]$ 
is defined by the following equation
\[ A_S=\int^{t_2}_{t_1} dt \int_{\partial W} ({\bf u},{\bf p}_n) dS_d . \]
The heat that are influx into the region $W$ through the surface $\partial W$
is defined by
\[ Q_S=\int^{t_2}_{t_1} dt \int_{\partial W} q_n dS_d ,\]
where $q_n=({\bf n},{\bf q})=n_k q_k$ is the density of heat flow.
Here, ${\bf n}$ is the vector of normal. 

The velocity of the total energy change is equal to the sum of power of 
mass force and the power of surface forces, and the energy flow 
from through the surface:
\be \label{E1}
\frac{d}{dt} \int_W \rho \Bigl( \frac{{\bf u}^2}{2}+e\Bigr) dV_D=
\int_W ({\bf u},{\bf f}) \rho dV_D+
\int_{\partial W}  ({\bf u},{\bf p}_n) dS_d+
\int_{\partial W} q_n dS_d .
\ee
Using Eq. (\ref{FTTD2}) for $A=\rho( {\bf u}^2/2+e)$, 
we can rewrite left-hand side of Eq. (\ref{E1}) in the form
\[ \frac{d}{dt} \int_W \rho \Bigl( \frac{{\bf u}^2}{2}+e\Bigr) dV_D=
\int_W \Bigl( \Bigl(\frac{d}{dt}\Bigr)_D 
\rho \Bigl( \frac{{\bf u}^2}{2}+e\Bigr)+
\rho \Bigl( \frac{{\bf u}^2}{2}+e\Bigr) \ Div_D {\bf u} \Bigr) dV_D = \]
\[ = \int_W\Bigl( 
\rho \Bigl(\frac{d}{dt}\Bigr)_D \Bigl( \frac{{\bf u}^2}{2}+e\Bigr)+
\Bigl(\Bigl(\frac{d}{dt}\Bigr)_D \rho+\rho \ Div_D{\bf u} \Bigr)
\Bigl( \frac{{\bf u}^2}{2}+e\Bigr) \Bigr)dV_D . \]
Using the equation of continuity, we get 
\[ \frac{d}{dt} \int_W \rho \Bigl( \frac{{\bf u}^2}{2}+e\Bigr) dV_D=
\int_W\Bigl( \rho \Bigl(\frac{d}{dt}\Bigr)_D 
\Bigl( \frac{{\bf u}^2}{2}+e\Bigr) \Bigr)dV_D = \]
\be \label{LHS}
 = \int_W\Bigl( \rho {\bf u} \Bigl(\frac{d}{dt}\Bigr)_D{\bf u}+ \rho 
\Bigl(\frac{d}{dt}\Bigr)_D e({\bf R},t) \Bigr)dV_D . \ee
The surface integrals in the right-hand side of Eq. (\ref{E1})
can be represented as  volume integrals
\be \label{RHS1} \int_{\partial W} ({\bf u},{\bf p}_n) dS_d=
\int_W \nabla^D_l ({\bf p}_l, {\bf u}) dV_D , \ee
and
\be \label{RHS2} \int_{\partial W} q_n dS_d=
\int_W \nabla^D_k q_k dV_D . \ee
Substituting Eqs. (\ref{LHS}) - (\ref{RHS2}) 
in Eq. (\ref{E1}), we get
\be \label{s3}
\int_W \Bigl( \rho {\bf u} \Bigl(\frac{d}{dt}\Bigr)_D{\bf u}+ \rho 
\Bigl(\frac{d}{dt}\Bigr)_D e({\bf R},t) \Bigr)dV_D =
\int_W  \Bigl( ({\bf u}, {\bf f}) \rho+
\nabla^D_l ({\bf p}_l, {\bf u}) + \nabla^D_k q_k \Bigr) dV_D . \ee
These equations can be rewritten in an equivalent form
\be \label{s3a}
\int_W \Bigl( \rho u_k \Bigl(\frac{d}{dt}\Bigr)_D u_k+ \rho 
\Bigl(\frac{d}{dt}\Bigr)_D e \Bigr)dV_D =
\int_W  \Bigl( \rho u_k f_k+
\nabla^D_l (p_{kl}u_k) + \nabla^D_k q_k \Bigr) dV_D . \ee
Let us use equations of balance of momentum (\ref{EM}). 
Multiplying both sides of these equations on the components $u_k$
of vector ${\bf u}$
and summing with respect to $k$ from 1 to 3, we get the equation
\be \label{EM2c} \rho u_k \Bigl(\frac{d}{dt}\Bigr)_D u_k =
\rho u_k f_k +u_k \nabla^D_l p_{kl}. \ee
Substituting Eq. (\ref{EM2c}) in Eq. (\ref{s3a}), we get
\be \label{s2}
\int_W \Bigl( \rho u_k f_k +u_k \nabla^D_l p_{kl} + \rho 
\Bigl(\frac{d}{dt}\Bigr)_D e \Bigr)dV_D =
\int_W  \Bigl( \rho u_k f_k+
\nabla^D_l ( p_{kl} u_k) + \nabla^D_k q_k \Bigr) dV_D . \ee
Using Eq. (\ref{7}) in the form
\[ \nabla^D_l (p_{kl} u_k) =
u_k\nabla^D_l p_{kl} +c(D,d,R) p_{kl} \frac{\partial u_k}{\partial x_l} , \] 
we obtain
\be \int_W \Bigl( \rho \Bigl(\frac{d}{dt}\Bigr)_D e - c(D,d,R)
p_{kl} \frac{\partial u_k}{\partial x_l}-
\nabla^D_k q_k \Bigr) dV_D=0 . \ee
This equation is satisfied for all regions $W$. Therefore, we get
the fractional equation of balance of density of energy in the form
\be \rho\Bigl(\frac{d}{dt}\Bigr)_D e= c(D,d,R)
p_{kl} \frac{\partial u_k}{\partial x_l}+\nabla^D_k q_k . \ee

%%%%%%%%%%%%%%%%%%%%%%%%%%%%%%%%%%%%%%%%%%%%%%%%%%%%%%%%
\section{Fractional Navier-Stokes and Euler Equations}

In Sections 3-5, we derive the fractional generalizations of the 
balance equations for fractal media. As the result we have 
the fractional hydrodynamic equations in the form: \\

(1) The fractional equation of continuity
\be \label{1eq} \Bigl(\frac{d}{dt}\Bigr)_D \rho=-\rho \nabla^D_k u_k . \ee

(2) The fractional equation of balance of density of momentum 
\be \label{2eq} \rho \Bigl(\frac{d}{dt}\Bigr)_D u_k=
\rho f_k+\nabla^D_l p_{kl} . \ee

(3) The fractional equation of balance of density of energy 
\be \label{3eq} \rho\Bigl(\frac{d}{dt}\Bigr)_D e= c(D,d,R)
p_{kl} \frac{\partial u_k}{\partial x_l}+\nabla^D_k q_k . \ee

\noindent
Here, we mean the sum on the repeated index $k$ and $l$ from 1 to 3.
We use the following notations
\be \nabla^D_k A= a(D,d) R^{3-D} \frac{\partial }{\partial x_k}  
\Bigl( R^{d-2} A \Bigr) . \ee
\be \Bigl(\frac{d}{dt}\Bigr)_D=\frac{\partial}{\partial t}+
c(D,d,R) u_l \frac{\partial}{\partial x_l} =
\frac{\partial}{\partial t}+
a(D,d) R^{d+1-D} u_l \frac{\partial}{\partial x_l} , \ee
where $a(D,d)$, and $c(D,d,R)$ are defined by the equations
\[ c(D,d,R)=a(D,d) R^{d+1-D}, \quad 
a(D,d)=\frac{2^{D-d-1} \Gamma(D/2)}{\Gamma(3/2) \Gamma(d/2)} , 
\quad R=\sqrt{\sum^3_{k=1} x^2_k} . \]
The equations of balance of density of mass, density of momentum and 
density of internal energy makes up a set of five equations,
which are not closed.
These equations, in addition to the hydrodynamic fields
$\rho({\bf R},t)$, $u({\bf R},t)$, $e({\bf R},t)$, include
also the tensor of viscous stress $p_{kl}({\bf R},t)$
and the vector of thermal flux $q_k({\bf R},t)$. 

Let us start with the definition of tensor $p_{kl}=p_{kl}({\bf R},t)$.
According to Newton's law, the force of viscous friction is 
proportional to the relative velocity of motion of medium layers
(that is to the gradient of the relevant component of velocity).
We further assume that tensor $p_{kl}({\bf R},t)$ is symmetrical,
and characterizes the dissipation due to viscous friction.
The most general form of tensor of viscous stress, which satisfies
the above requirements, is determined by two constants ($\mu$ and $\xi$) 
which can be chosen so that
\be \label{pkl} p_{kl}=
-p\delta_{kl} +\mu \Bigl(
\frac{\partial u_k}{\partial x_l}+\frac{\partial u_l}{\partial x_k}
-\frac{2}{3}\delta_{kl}\frac{\partial u_m}{\partial x_m}
\Bigr)+\xi \delta_{kl} \frac{\partial u_m}{\partial x_m} . \ee
This expression contains a second coefficient of viscosity $\xi$,
called the coefficient of internal viscosity because it reflects
the existence of internal structure of particles.
In case of structureless particles $\xi=0$. 

The definition of the vector of heat flux $q_k=q_k({\bf R},t)$,
is based on the empirical Fourier law
\be \label{FL} q_k=-\lambda \frac{\partial T}{\partial x_k}, \ee
where $T=T({\bf R},t)$ is the field of temperature.
The value of heat conductivity $\lambda$ can be found experimentally.

Now we have a closed set of Eqs. 
(\ref{1eq})-(\ref{3eq}), (\ref{pkl}), and (\ref{FL}) for fields
$\rho({\bf R},t)$, $u_k({\bf R},t)$, $T({\bf R},t)$ -
a set of fractional equations of hydrodynamics.
The fractional generalization of the equations of theory of 
elasticity for solids can be obtained in a similar way. 

Let us consider the special cases of the set of Eqs.
(\ref{1eq})-(\ref{3eq}), (\ref{pkl}), and (\ref{FL}). 

{\bf 1}. Let us consider the fluids that are defined by 
\[ p_{kl}=- p \delta_{kl} , \quad q_k=0 , \]
where $p=p({\bf R},t)$ is the pressure. 
The fractional hydrodynamic equations for these fluids 
have the following form
\be \label{Ee1} \Bigl(\frac{d}{dt}\Bigr)_D \rho=-\rho \nabla^D_k u_k . \ee
\be \label{Ee2} \Bigl(\frac{d}{dt}\Bigr)_D u_k=
f_k- \frac{1}{\rho}\nabla^D_k p . \ee
\be \label{Ee3} \Bigl(\frac{d}{dt}\Bigr)_D e= - c(D,d,R) \frac{p}{\rho}
\frac{\partial u_k}{\partial x_k}  . \ee
These equations are the fractional generalization of the Euler equations.

{\bf 2}. Let the coefficients $\mu$, $\xi$ and $\lambda$ are constants.
If we consider homogeneous viscous fluid, then we have
\[ \Bigl(\frac{d}{dt}\Bigr)_D\rho=0.  \]
For the fractal media, we have the non-solenoidal field of the velocity 
($div ({\bf u})\not=0$), that satisfies
the relation
\be \label{div0} 
div({\bf u})=\frac{\partial u_m}{\partial x_m}=
\frac{(2-d) x_k u_k}{\sum^3_{=1} x^2_l}. \ee

Using $q_k=0$ and Eq. (\ref{pkl}), 
we get the fractional generalization of 
Navier-Stokes equations in the form
\be \label{NS} \rho \Bigl( \frac{\partial u_k}{\partial t}+
c(D,d,R) u_l \frac{\partial u_k}{\partial x_l}\Bigr) =
\rho f_k- \nabla^D_k p +
\mu \ \nabla^D_l \frac{\partial u_k}{\partial x_l} +
\mu \ \nabla^D_l \frac{\partial u_l}{\partial x_k} +
(\xi-\frac{2}{3}\mu) \nabla^D_k \frac{\partial u_l}{\partial x_l} . \ee
Equations (\ref{div0}) and (\ref{NS}) form the system of 4 equations for
4 fields $u_1({\bf R},t),u_2({\bf R},t),u_3({\bf R},t)$, and $p({\bf R},t)$.  
Note that Eq. (\ref{NS}) can be rewritten in an equivalent form
\[ \rho \Bigl( \frac{\partial u_k}{\partial t}+
c(D,d,R) u_l \frac{\partial u_k}{\partial x_l}\Bigr) = \]
\be \label{NS2} =\rho f_k- c(D,d,R)\frac{\partial p}{\partial x_k}+
\mu \  c(D,d,R) \frac{\partial^2 u_k}{\partial x_l \partial x_l} +
(\xi+\frac{\mu}{3}) c(D,d,R) \frac{\partial^2 u_l}{\partial x_k \partial x_l}
+ L_k(D,d,R,u) . \ee
Here, we use the following notations:
\[ L_k(D,d,R,u)= \mu \Bigl(
\frac{\partial u_k}{\partial x_l}+\frac{\partial u_l}{\partial x_k}
\Bigr) \nabla^D_l(1) -p  \nabla^D_k(1)
+(\xi-\frac{2}{3}\mu) \frac{\partial u_l}{\partial x_l} \nabla^D_k(1) , \]
where $\nabla^D_k(1)=c^{-1}_3(D,R) \partial c_2(d,R)/ \partial x_k$. 
If $c(D,d,R)=1$ and $ L_k(D,d,R,u)=0$, then Eq. (\ref{NS2}) has the usual
form of the Navier-Stokes equations.

%%%%%%%%%%%%%%%%%%%%%%%%%%%%%%%%%%%%%%%%%%%%%%%%%%%%%%%%%%%%
\section{Fractional Equilibrium Equation}

The equilibrium state of media means that we have the conditions
\[ \frac{\partial A}{\partial t}=0, \quad \frac{\partial A}{\partial x_k}=0 , \]
for the hydrodynamic fields $A=\{\rho, u_k, e \}$.
In this case, the fractional hydrodynamic equations have the form
\[ \rho f_k+\nabla^D_l p_{kl}=0, \quad \nabla^D_k q_k=0. \]
Using $\partial u_l/ \partial x_k=0$, 
we get that the tensor $p_{kl}$ has the form
$p_{kl}=-p\delta_{kl}$.
Using the Fourier law, we have the following system of equations: 
\be \label{ee1}f_k=\frac{1}{\rho}\nabla^D_k p, \ee
\be \label{ee2} \nabla^D_k(\lambda \frac{\partial T}{\partial x_k})=0 , \ee
which are the fractional generalization of the equilibrium equations 
for the media.
Eq. (\ref{ee1}) can be rewritten in an equivalent form
\[ \frac{\partial (c_2(d,R) p)}{\partial x_k}=\rho c_3(D,R) f_k . \]
Let us consider the homogeneous medium with the density
$\rho(x)=const$.
In this case, we have the equation
\[ c_3(D,R)f_k=\frac{\partial (c_2(d,R)p/\rho_0) }{\partial x_k} . \]
If the force $f_k$ is a potential force such that 
$c_3(D,R)f_k=-\partial U/ \partial x_k$, 
then we get the fractional generalization of equilibrium equation
in the form
\be \label{eqeq1} c_2(d,R)p+\rho_0 U =const. \ee

%%%%%%%%%%%%%%%%%%%%%%%%%%%%%%%%%%%%%%%%%%%
\section{Fractional Bernoulli Integral}

Let us consider the equation of balance of momentum density
with the tensor $p_{kl}=-p\delta_{kl}$. Using the relation
\[ \Bigl(\frac{d}{dt}\Bigr)_D \frac{{\bf u}^2}{2}=
u_k \Bigl(\frac{d}{dt}\Bigr)_D u_k , \]
and Eq. (\ref{2eq}), we get 
\be \label{Bi1} \Bigl(\frac{d}{dt}\Bigr)_D \frac{{\bf u}^2}{2}=
u_k f_k-\frac{1}{\rho} u_k \nabla^D_k p . \ee
If the potential energy $U$ and pressure $p$ is 
time-independent fields (${\partial U}/{\partial t}=0$, 
${\partial p}/{\partial t}=0$), then we can use 
the following relation:
\be \label{d/dt} \Bigl(\frac{d}{dt}\Bigr)_D = c(D,d,R) \frac{d}{dt} . \ee
Let us consider the non-potential force that is described by the equation
\be \label{npf} f_k=-c(D,d,R)\partial U/ \partial x_k . \ee 
If $D=3$ and $d=2$, then this force is potential. 
Using Eqs. (\ref{d/dt}) and (\ref{npf}), we can rewrite
Eq. (\ref{Bi1}) in the form
\[ \frac{d}{dt}\Bigl( \frac{{\bf u}^2}{2}+U(D,d)+P(d) \Bigr)=0 , \]
where the function $P$ is defined by the usual relation
\[ P(d)=\int^p_{p_0} \frac{d(c_2(d,R)p)}{c_2(d,R)\rho} .  \]
As the result we have that the integral
\[ \sum^3_{k=1}\frac{u^2_k}{2}+U+P(d) =const . \]
can be considered as a fractional generalization
of Bernoulli integral for fractal media. 
If the forces $f_k$ are potential, then the fractional analog 
of the Bernoulli integral does not exists.
If the density is described by
\be \label{rhoc2}
\rho=\rho_0 c^{-1}_2(d,R)=\rho_0 \frac{\Gamma(d/2)}{2^{2-d}} R^{2-d} , \ee
then we have the fractional generalization of Bernoulli integral 
in the following form
\[ \frac{\rho_0 {\bf u}^2}{2}+\rho_0 U(D,d)+ c_2(d,R) p = const . \]
If $u_k=0$, then we get equilibrium Eq. (\ref{eqeq1}) for 
nonpotential force (\ref{npf}) and the density (\ref{rhoc2}).

%%%%%%%%%%%%%%%%%%%%%%%%%%%%%%%%%%%%%%%
\section{Sound Waves in Fractal Media}

Let us consider the motion of the medium with small perturbations.
Fractional equations of motion (\ref{Ee1}) and (\ref{Ee2}) have the form
\be \label{em1} \frac{\partial \rho}{\partial t}+
c(D,d,R) u_l \frac{\partial \rho}{\partial x_l}=-\rho \nabla^D_k u_k , \ee
\be \label{em2} \frac{\partial u_k}{\partial t}+
c(D,d,R) u_l \frac{\partial u_k}{\partial x_l}=
f_k- \frac{1}{\rho}\nabla^D_k p . \ee
Let us consider the small perturbation
\be \label{sp} \rho=\rho_0+\rho^{\prime}, \quad p=p_0+p^{\prime}, \quad 
u_k=u^{\prime}_k, \ee
where $\rho^{\prime}<< \rho_0$, and $p^{\prime}<< p_0$.
Here, $p_0$ and $\rho_0$ describe the steady state that is defined by
the conditions
\[ \frac{\partial \rho_0}{\partial t}=0,\quad
\frac{\partial \rho_0}{\partial x_k}=0, \quad
\frac{\partial p_0}{\partial t}=0, \quad
\frac{\partial p_0}{\partial x_k}=0 . \]
Suppose that $f_k=0$. 
Substituting (\ref{sp}) in Eqs. (\ref{em1}) and (\ref{em2}),
we derive the following equations
for the first order of the perturbation: 
\be \label{em3} \frac{\partial \rho^{\prime}}{\partial t}
=-\rho_0 \nabla^D_k u^{\prime}_k , \ee
\be \label{em4} \frac{\partial u^{\prime}_k}{\partial t}=
- \frac{1}{\rho_0}\nabla^D_k p^{\prime} . \ee
These equations are equations of motion for the small perturbations.
To derive the independent equations for perturbations,  
we consider the partial derivative of Eq. (\ref{em3})
with respect to time:
\be \label{em5} \frac{\partial^2 \rho^{\prime}}{\partial t^2}
=-\rho \nabla^D_k \frac{ \partial u^{\prime}_k}{\partial t} . \ee
Substituting Eq. (\ref{em4}) in this equation, we get
\be \label{em6} \frac{\partial^2 \rho^{\prime}}{\partial t^2}
=\nabla^D_k \nabla^D_k p^{\prime} . \ee
If we consider the adiabatic processes, we can use equation $p=p(\rho,s)$. 
For the first order of perturbation we have the equation
\[ p^{\prime}=v^2 \rho^{\prime} , \quad 
v=\sqrt{\Bigl(\frac{\partial p}{\partial \rho} \Bigr)_s} . \]
As the result, we obtain the following 
fractional generalization of the wave equations: 
\be \frac{\partial^2 \rho^{\prime}}{\partial t^2}-
v^2 \nabla^D_k \nabla^D_k \rho^{\prime}=0 , \ee
\be  \frac{\partial^2 p^{\prime}}{\partial t^2}-
v^2 \nabla^D_k \nabla^D_k p^{\prime}=0 . \ee

Let us consider the simple example of the fractional wave equations. 
If we consider one dimensional case ($n=1$), where $D<1$ and $c_2=1$,
then we get the following equation for pressure 
\be c_1(D,x)\frac{\partial^2 p^{\prime}}{\partial t^2}-
v^2 \frac{\partial}{\partial x} \Bigl( c_1(D,x)
\frac{\partial p^{\prime}}{\partial x} \Bigr)=0 ,
\ee
where the coefficient $c_1(D,x)$ is defined by the relation
\[ c_1(D,x)=\frac{|x|^{D-1}}{\Gamma(D)} . \]

%%%%%%%%%%%%%%%%%%%%%%%%%%%%%%%%%%%%%%%%%%%%%%%%%%%%%%
\section{Solution of the Fractional Wave Equation}

The fractional generalization of wave equation has the following form
\be \label{sp2} c_1(D,x) \frac{\partial^2 u}{\partial t^2}=
\frac{\partial u}{\partial x}
\Bigl( v^2 c_1(D,x) \frac{\partial u}{\partial x}\Bigr) , \ee
where $c_1(D,x) \ge 0$.
Let us consider the region $0\le x \le l$ and the following conditions:
\[ u(x,0)=f(x), \quad (\partial_t u)(x,0)=g(x), \]
\[ u(0,t)=0, \quad u(l,t)=0 . \]
The solution of Eq. (\ref{sp2}) has the form
\[ u(x,t)=\sum^{\infty}_{n=1} \Bigl( f_n cos(\lambda_n t)+
\frac{g_n}{\sqrt{\lambda_n}} sin(\lambda_n t) \Bigr) y_n(x) . \]
Here, $f_n$ and $g_n$ are the Fourier coefficients for the functions
$f(x)$ and $g(x)$ that are defined by the equations
\[ f_n=||y_n||^{-2} \int^l_0 f(x) y_n(x) d l_D =
||y_n||^{-2} \int^l_0 c_1(D,x) f(x) y_n(x) d x, \]
\[ g_n=||y_n||^{-2} \int^l_0 f(x) y_n(x) d l_D=
||y_n||^{-2} \int^l_0 c_1(D,x) f(x) y_n(x) dx, \]
\[ ||y_n||^2=\int^l_0 y^2_n(x) d l_D =\int^l_0 c_1(D,x) y^2(x) dx , \]
where $dl_D=c_1(D,x)dl_1$, and $dl_1=dx$.
Note that the eigenfunctions $y_n(x)$ satisfy the following condition
\[ \int^l_0 y_n(x) y_m(x) d l_D=\delta_{nm} . \]
The eigenvalues $\lambda_n$ and the eigenfunctions $y_n(x)$
are defined as solutions of the equation
\[ v^2[c_1(D,x)y^{\prime}_x]^{\prime}_x+\lambda^2 c_1(D,x) y=0, \quad
y(0)=0, \quad y(l)=0 . \] 
This equation can be rewritten in an equivalent form
\[ v^2 xy^{\prime \prime}_{xx}(x)+(D-1)y^{\prime}_{x}(x)+ \lambda^2 x y(x)=0 . \]
The solution of this equation has the form
\[ y(x)=C_1x^{1-D/2}J_{\nu}(\lambda x/v)+
C_2 x^{1-D/2} Y_{\nu}(\lambda x/v) , \] 
where $\nu= |1-D/2|$. 
Here, $J_{\nu}(x)$ are the Bessel functions of the first kind,
and $Y_{\nu}(x)$ are the Bessel functions of the second kind.

As an example, we consider the case that is defined by
\[ l=1, \quad v=1, \quad 0\le x \le 1, \quad f(x)=x(1-x), \quad g(x)=0 . \]
The usual wave has $D=1$ and the solution
\[ u(x,t)=\sum^{\infty}_{n=1} \frac{4(1- (-1)^n) 
sin(\pi n x) cos (\pi n t) }{\pi^3 n^3} . \]
The approximate solution for the usual wave with $D=1$
that has the form
\[ u(x,t) \simeq \sum^{10}_{n=1} \frac{4(1- (-1)^n) 
sin(\pi n x) cos (\pi n t) }{\pi^3 n^3}  \]
is shown in Fig. 1 for $0\le t \le 3$ and velocity $v=1$.  

%%%\begin{figure}[tbh]
%%%\begin{center}
%%%\resizebox{17cm}{!}{\includegraphics{US10d.eps}}
%%%\caption{\it Usual Wave (D=1) with the velocity $v=1$.}  
%%%\label{US10}
%%%\end{center}
%%%\end{figure}

%%%\begin{figure}[tbh]
%%%\begin{center}
%%%\resizebox{17cm}{!}{\includegraphics{FS10d.eps}}
%%%\caption{\it Fractal Media Wave (D=1/2) with the velocity $v=1$.}  
%%%\label{FS10}
%%%\end{center}
%%%\end{figure}

If $D=1/2$, then we have the fractal medium wave with
\[ y_n(x)=\frac{1}{\Gamma(1/2)} x^{3/4} J_{3/4} (\sqrt{2} \lambda_n x/2) . \]
The eigenvalues $\lambda_n$ are the zeros of the Bessel function
\[ \lambda_n: \quad J_{3/4} (\sqrt{2} \lambda_n /2)=0 . \] 
For example,  
\[ \lambda_1 \simeq 4.937, \quad \lambda_2 \simeq 9.482, \quad
 \lambda_3 \simeq 13.862, \quad \lambda_4 \simeq 18.310, \quad
 \lambda_5 \simeq 22.756 . \]
The approximate values of the eigenfunctions
\[ f_n=\frac{||y_n||^{-2}}{\Gamma(D)} \int^l_0 x^{5/4} (1-x) 
J_{3/4} (\sqrt{2} \lambda_n /2) dx  , \]
are following
\[ f_1 \simeq 1.376, \quad f_2 \simeq -0.451, \quad f_3 \simeq 0.416, \quad
f_4 \simeq -0.248, \quad f_5 \simeq 0.243. \]
The solution of the fractional equation
\[ u(x,t)=\sum^{\infty}_{n=1} f_n cos(\lambda_n t) 
J_{3/4}(\sqrt{2} \lambda_n x /2) .\]
The approximate solution for the fractal media wave with $D=1/2$ that
has the form
\[ u(x,t) \simeq 
\sum^{10}_{n=1} f_n cos(\lambda_n t) J_{3/4}(\sqrt{2} \lambda_n x /2)  \]
is shown in Fig. 2 for the velocity $v=1$.

%\newpage
%%%%%%%%%%%%%%%%%%%%%%%%%%%%%%%%%%%%%%%%%%%%%%%%%%%%%%%%%%%%%%%
\section{Conclusion}

The fractional continuous models of fractal media
can have a wide application. 
This is due in part to the relatively small numbers of parameters 
that define a random fractal medium of great complexity
and rich structure. The fractional continuous model allows us
to describe dynamics for wide class fractal media.  

In many problems the real fractal structure of matter 
can be disregarded and the medium can be replaced by  
some "fractional" continuous mathematical model. 
To describe the medium with 
non-integer mass dimension, we must use the fractional calculus.
Smoothing of the microscopic characteristics over the 
physically infinitesimal volume transform the initial 
fractal medium into "fractional" continuous model
that uses the fractional integrals. 
The order of fractional integral is equal 
to the fractal mass dimension of the medium.

The experimental research of the hydrodynamics of fractal media
can be realized by introducing a neutral indicator,
its distribution. It allows one to obtain much information
on the motion and mixing of fluid. 
The fractal nature of damage and porosity 
has been experimentally detected over a wide range of scales.

Note that the fractional hydrodynamic equations for fractal media
can be derived from the fractional generalization of the Bogoliubov 
equations that are suggested in \cite{PRE05}. \\

%%%\section*{Acknowledgment}

I would like to thank 
Prof. G.M.  Zaslavsky for very useful discussions. \\

%%%%%%%%%%%%%%%%%%%%%
\section*{Appendix: Fractional Gauss Theorem}

To realize the representation, we derive the fractional 
generalization of the Gauss theorem
\be \label{GT}
\int_{\partial W} A u_n dS_2 =\int_W div( A {\bf u}) dV_3 ,
\ee
where $u_n$ is defined by $u_n=({\bf u},{\bf n})=u_kn_k$, 
the vector ${\bf u}=u_k{\bf e}_k$ is a velocity field, 
and ${\bf n}=n_k {\bf e}_k$ is a vector of normal. Here,  
\[ div( A {\bf u})=\frac{\partial (A {\bf u})}{\partial {\bf R}}= 
\frac{\partial (A u_k)}{\partial x_k} . \]
Here, and later we mean the sum on the repeated index
$k$ and $l$ from 1 to 3.

Using the relation
\be \label{c2} dS_d=c_2 (d,R)dS_2 , \quad 
c_2(d,R)= \frac{2^{2-d}}{\Gamma(d/2)} |{\bf R}|^{d-2} , \ee
we get
\[ \int_{\partial W} A u_n dS_d =\int_{\partial W}  c_2(d,R) A u_n dS_2 . \]
Note that we have $c_2(2,R)=1$ for the $d=2$. 
Using the usual Gauss theorem (\ref{GT}), we get 
\[ \int_{\partial W}  c_2(d,R) A u_n dS_2 =\int_W  div(c_2(d,R) A {\bf u}) dV_3 . \]
The relation
\be \label{c3} dV_D=c_3 (D,R)dV_3 , \quad 
c_3(D,R)= \frac{2^{3-D} \Gamma(3/2)}{\Gamma(D/2)} |{\bf R}|^{D-3}  \ee
in the form $dV_3=c^{-1}_3(D,R) dV_D$
allows us to derive the fractional generalization of the Gauss theorem:
\be \label{FGT}
\int_{\partial W} A u_n dS_d =\int_W c^{-1}_3(D,R) div( c_2(d,R) A {\bf u}) dV_D .
\ee

%\newpage
%%%%%%%%%%%%%%%%%%%%%%%%%%%%%%%%%%%%%%%%%%%%%%%%%%%%%%%%%%%

\end{document}